\documentclass[12pt]{article}
\newcommand\ch{\mathop{\rm ch}\nolimits}
\newcommand\sh{\mathop{\rm sh}\nolimits}

\newcommand\sgn{\mathop{\rm sgn}\nolimits}

\begin{document}

\title{$\phi$, $\Omega$ and $\rho$ 
production from deconfined matter \\
       in relativistic heavy ion collisions at CERN SPS}
\author{ P. Csizmadia$^{1}$,  P. L\'evai$^{1,2}$\\[1ex]
$^{1}$ RMKI Research Institute for Particle and Nuclear Physics, \\
\ \  P. O. Box 49, Budapest, 1525, Hungary \\
$^{2}$ Dept. of Physics, Columbia University,
 New York, NY, 10027, USA
 }

\date{September 28, 1999}

\maketitle
\begin{abstract}
We investigate the production of the $\phi$ meson and the $\Omega$ baryon 
which interact weakly with hot hadronic matter, and thus their spectra
reflect the early stage of the heavy ion collisions.
Our analysis shows that the hadronization temperature, $T_{had}$,
and the transverse flow, $v_T^0$, of the initial deconfined phase 
are strongly correlated:
$T_{had} + a \cdot (v_T^0)^2=0.25$ GeV, where $a=0.37$ GeV
in the Pb+Pb collision at 158 AGeV/c. 
When choosing appropriate initial values of $T_{had}$ and $v_T^0$ from the
temperature region $T_{had}=175 \pm 15$ MeV, the measured $\rho$ meson spectra
was  reproduced surprisingly well by the MICOR model. 
We have found weak influence of final state hadronic 
interactions on the transverse hadron spectra at $m_T - m_i > 0.3$GeV.
\end{abstract}

Recent analysis of NA50 Collaboration \cite{QM99NA50}
on $\phi$ and $\rho/\omega$ meson transverse momentum spectra
in Pb+Pb collision at the CERN SPS, yielded remarkably
small values for the inverse slopes of these particles, namely
$T(\phi)=222 \pm 6$ MeV and $T(\rho/\omega)=219 \pm 5$ MeV \cite{K1Bes}. 
Furthermore, the analysis by the
WA97 Collaboration \cite{SQM98WA97} reveals  a slightly larger
value for the inverse slope of the $\Omega^-+\overline{\Omega}^+$
particle spectra, $T(\Omega)=238 \pm 17$ MeV \cite{Ex32}.

These inverse slopes do not fit the suggested straight line
of $T_i=T_{fo} + m_i \langle v_T \rangle^2$ \cite{CsorLor,NA44prl},
which may indicate a mutual hadronic transverse flow and which holds
for $\pi^\pm$, $K^\pm$, $p$, $\overline p$, $\phi$, $\Lambda$,
$\overline{\Lambda}$ and $d$  data of 
NA44 \cite{NA44prl} and NA49 \cite{NA49phi,Roland98,Grabler99} 
with freeze-out temperature $T_{fo}=175$ MeV and average
transverse flow $\langle v_T \rangle =0.35c$.
On the other hand, the measured slopes of
the $\Xi^-$ and ${\overline \Xi}^+$  did not follow  
this linear dependence on hadron masses $m_i$ \cite{SQM98WA97}.
This phenomena was explained by the early freeze-out
of the $\Xi$ and the $\Omega$ in an analysis
based on the RQMD model \cite{HecSorge98}, 
where final state interactions were considered to reproduce 
the measured hadronic transverse slopes. 
Since the interaction of the $\phi$ meson with non-strange baryon-rich
matter is also small \cite{Phi}, the measured 
value of $T(\phi)$ from NA50 agrees with this picture.

However, accepting that the momentum spectra of the $\phi$ and $\Omega$ 
reflect the early stage of the heavy ion collisions,  
we can ask if this early stage could have been a deconfined matter, or
simply resonance matter as indicated by RQMD \cite{HecSorge98}.

In this paper we investigate the direct production of 
the $\Omega$, $\phi$ and $\rho$ from 
thermalized massive quark matter. We use the
MIcroscopical COalescence Rehadronization (MICOR) model
\cite{Hirschegg97,MICORSQM98,MICORUDQ99},
which is the successor of the ALCOR \cite{ALCOR1,ALCOR2}
and the Transchemistry models \cite{Transchem}. The ALCOR and Transchemistry
models were constructed to determine the total number of
hadrons produced from quark matter. 
The MICOR model is able to determine
the full momentum spectra of final state hadrons.
For the investigation of
the production of $\phi$ and $\Omega$ we use a part of the 
full MICOR model, which part will be summarized here.
A full scale description of the MICOR model and the detailed
investigation of all particle spectra will be published elsewhere 
\cite{MICOR}.

One description of the deconfined phase consists of massive quarks,
antiquarks and gluons in the temperature region
$1 \leq T/T_c \leq 3$ \cite{LevHein}.
In the MICOR model, which is based on this description, the massive gluons
(whose number is suppressed relative to the quarks and antiquarks)
become responsible for an attractive effective
potential between the colorful quarks and anti-quarks and
this effective strong interaction drives the hadronization via
the coalescence of quarks. The obtained quark-antiquark plasma is assumed 
to be a fully thermalized state, and in the hadronization stage
it is characterized by a hadronization temperature $T_{had}$,
a transverse flow $v_T^0$ and a Bjorken-scaled longitudinal flow.
In spite of the thermal initial conditions,
the coalescence of the massive quarks and antiquarks produces a
hadron resonance gas which is out of equilibrium. Furthermore,  the decays
of the resonances do not lead to equilibrium, but keep the
final state hadrons in their out of equilibrium state.
While the final state hadronic spectra can be fitted by inverse slopes
familiar from equilibrium descriptions,  this does not mean that
we have equilibrated hadronic final states in our calculation.

During hadronization meson-like objects are formed in one step 
via quark-antiquark coalescence.  Baryon-like objects are formed 
in two steps: the formation of diquarks from quarks are followed by
the coalescence of diquarks and quarks into baryons.
The presence of strongly correlated hadron-like objects
in the deconfined phase is supported by lattice-QCD results
\cite{CorrLatt}.
We use the name {\it 'prehadron'} for such a correlator.
The values of prehadron masses, $m_{preh}$, are not sharply determined.
In the static case one may determine the spectral function of the correlator,
extracting some 'mass' and 'width'. However, in our case
the two-body coalescence process dynamically creates an off-shell prehadron
with $\sigma_{coal}$ cross section
(see Ref.~\cite{ALCOR1} for the coalescence cross section). 

We assume that during the phase transition period
these color-neutral prehadrons  can escape the deconfined state
without disintegration and they will
become the species of the produced excited hadronic gas.
Large constituent quark masses imply a
further assumption that the production of
the excited J=1 vector mesons ($\rho,K^*,\phi$)
and J=3/2 baryons ($\Delta$, $\Sigma^*$, $\Xi^*$, $\Omega$) are favored
during the hadronization.  The J=0 pseudoscalar
mesons (pions and Kaons) can not be formed by quark coalescence, 
but appear together with the J=1/2 baryons
as products of the decay of the heavier excited hadrons.
The expected minor difference between the prehadron masses and the
excited hadron masses will be corrected in a last step:
when the prehadron escapes from the deconfined region, it becomes on-shell
by conserving its velocity.

In general, the coalescence-type hadron production, 
$q_1 + q_2 \longrightarrow h$,  can be described by a relativistic
rate equation based on the densities \cite{hadrochemical-rateeq}:
\begin{eqnarray}
\partial_\mu(n_h u^\mu)\ =\ \sum_{q_1,q_2}
\langle\sigma_{q_1 q_2}^h v_{12}\rangle n_{q_1} n_{q_2} \ .
\label{rateq.hadrochem}
\end{eqnarray}
Here $n_{q_1}$ and $n_{q_2}$ are the local quark densities,
$n_h$ is the prehadron density,
$u^\mu$ is the four-velocity of the matter,
$v_{12}$ is the relative velocity of the two quarks, 
$\sigma_{q_1 q_2}^h$ is the quark coalescence cross section and
$\langle\sigma_{q_1 q_2}^h v_{12}\rangle$ 
is the momentum space average of their
product:
\begin{equation}
\langle\sigma_{q_1  q_2}^h v_{12}\rangle\ =\ {\int d^3p_1 d^3p_2 f_1 f_2
	\sigma_{q_1 q_2}^h v_{12} \over \int d^3p_1 d^3p_2 f_1 f_2}.
\label{s_averaging}
\end{equation}
For the invariant quark distributions, $f_i(p_i, x_i)$,
thermalized J\"uttner-functions are used. However,
in general, other momentum distributions
could also be used. We neglect the melting of the prehadrons 
back into the deconfined phase. This equation
assumes the escape of color-neutral prehadrons from the deconfined phase.

In the MICOR model we determine the momentum spectra
of the produced prehadrons
generalizing a momentum dependent version of the rate
equation found in eq.~(\ref{rateq.hadrochem}). We introduce an
extended averaging on the $\tau=\tau_h$ hadronization hypersurface:
\begin{equation}
\langle\langle\sigma v_{12}\rangle\rangle
\ =\ {{\hat I [f_1 f_2\sigma v_{12}]}\over {\hat I [f_1 f_2]}} \ ,
   \label{rateq.inhom.coalfactor}
   \end{equation}
   where
   \begin{eqnarray}
   \hat I [...] &=& \int dV_1 dV_2 d^3p_1 d^3p_2\  [...] = \\
	  &=& \int dV_1 dV_2 d^4p_1 d^4p_2\Theta(E_1)\Theta(E_2)
     (4E_1E_2)\delta(p_1^2-m_1^2)\delta(p_2^2-m_2^2) \ [...]
     \label{rateq.inh2}
     \end{eqnarray}
     is the 12 dimensional phase space integration operator.
 Assuming energy-momentum conservation, the momentum of the outgoing particle
 is the sum of the incoming momenta: $p=p_1+p_2$.
 Since we are interested in the momentum spectra of the outgoing particle, 
 it is more useful to represent $\hat I$ in
 eq.~(\ref{rateq.inh2}) as an integral of the
 relative four-momentum $q=(p_1-p_2)/2$ and the outgoing four-momentum $p$ 
 with $s=p^2$:
 \begin{eqnarray}
 \hat I[...] &=& {\int d^4p \hat I_p}[...] \ ,\\
 \hat I_p[...]
 &=& \intop_{\tau=\tau_h}\!\! dV_1 dV_2\!\!\!\!\!\sum_{\sgn q^0=\pm 1}
 \!\!\!\!\!\!\!\!\intop_{q^2={m_1^2+m_2^2\over 2}-{s\over 4}}
 \!\!\!\!\!\!\!\!\!\!{d^3q\over |q^0|}
  E_1 E_2\delta\left(pq-{m_1^2-m_2^2\over 2}\right)
  \Theta\left({E\over 2}-|q^0|\right) [...] \ ,
\end{eqnarray}
Using the parametrization
\begin{eqnarray}
p^\mu&=&(m_T'\ch y,\ p_T'\cos\varphi,\ p_T'\sin\varphi,\ m_T'\sh y),\\
q^\mu&=&\pm\sqrt{|q^2|}(q_c,
\ q_s\sin\zeta\cos(\varphi+\chi),\ q_s\sin\zeta\sin(\varphi+\chi),
\ q_s\cos\zeta),
\end{eqnarray}
after some tedious algebra we obtain a simplified expression for
the phase space integral:
\begin{eqnarray}
\hat I_p [...] &=&\int dV_1 dV_2\sqrt{|q^2|}\sum_\pm
  \intop_{\Theta(q^2)}^{E\over 2\sqrt{|q^2|}}\!\!\!\! q_s dq_c
 \intop_{-1}^{+1} d\cos\zeta
2E_1 E_2{\Theta(X_\pm(\cos\zeta))\over\sqrt{X_\pm(\cos\zeta)}} \ [...] \ .
 \label{rateq.inhom.Ip}
\end{eqnarray}
Here
\begin{eqnarray}
X_\pm(\cos\zeta)&=&-a\cos^2\zeta+b_\pm \cos\zeta-c_\pm,\\
a&=&(E^2-s)q_s^2,\\
b_\pm&=&2\left(E q_c\mp {m_1^2-m_2^2\over 2\sqrt{|q^2|}}\right) p_z q_s,\\
c_\pm&=&\left(E q_c\mp {m_1^2-m_2^2\over 2\sqrt{|q^2|}}\right)^2 - p_T^2 q_s^2.
\end{eqnarray}
Thus, from eq.~(\ref{rateq.inhom.Ip}),
we derived the differential form of
the averaging in eq.~(\ref{rateq.inhom.coalfactor}):
\begin{equation}
{d\langle\langle\sigma v_{12}\rangle\rangle\over d^4p}
\ =\ {{\hat I_p [f_1 f_2\sigma v_{12}]}\over{\hat I [f_1 f_2]}} \ .
\label{rateq.inhom.4momdistr}
\end{equation}
The eq.~(\ref{rateq.inhom.4momdistr})
leads to a momentum dependent rate equation for the produced
prehadrons and diquarks.

In eq.~(\ref{rateq.inhom.4momdistr})
the prehadron  is created off-shell, $s=p^2=m_{preh}^2$.
When the prehadron leaves the deconfined region, it assumes to be an
on-shell resonance by emitting or absorbing energy.  
For simplicity, we assume that 
its velocity distribution remains unchanged during this process.
One can parametrize the transverse velocity as
\begin{equation}
\ch\mu\ =\ m_T'/\sqrt{s} \ ,
\end{equation}
where $m_T'=\sqrt{p_T'^2 +s}$.
With this parametrization the differential momenta element is
\begin{equation}
d^4p\ =\ s\, ds\, \ch\mu\, d\ch\mu\, dy\, d\varphi,
\end{equation}
and thus the four-velocity distribution is
\begin{equation}
{dI\over\ch\mu\, d\ch\mu\, dy\, d\varphi}
\ =\ \int ds\, s {d\langle\langle\sigma v_{12}\rangle\rangle\over d^4p} \ .
\label{didmu}
\end{equation}

If we substitute $\ch\mu=m_T/m_{h}$ into eq.~(\ref{didmu})
and convert the velocity distribution into a four-momentum
distribution, we obtain the on-shell
momentum distribution for hadrons with mass $m_h$ and transverse mass
$m_T=\sqrt{p_T^2+m_h^2}$:
\begin{equation}
{dI\over m_T dm_T dy\, d\varphi}
\ =\ {1\over m_h^2}\int ds\, s
     {d\langle\langle\sigma v_{12}\rangle\rangle\over d^4p} \ .
\label{rateq.inhom.momdistr}
\end{equation}
This formula gives the momentum distribution
of the primarily produced excited hadrons. 
From the expression
in eq.~(\ref{rateq.inhom.momdistr}) one can determine in one step
the transverse momentum slope of the $\phi$ and in two steps 
the $\Omega$ (included the formation of an $ss$ diquark). 
These results can be compared directly with the experimental data.
For the  description of all final state hadrons
we would need to follow the time-evolution of the multicomponent hadronization
and the decay of excited hadrons. This is beyond the scope of this 
paper and those calculations will be published in a following paper
\cite{MICOR}.

The deconfined  matter is characterized by the hadronization temperature, 
$T_{had}$, and an initial transverse flow, $v_T^0$.
We consider a large enough
longitudinal extension for the deconfined matter: $\eta_0 = \pm 1.8$,
where $\eta_0$ is the space-time rapidity.
This parameter will not influence the transverse momentum spectra at $y=0$.
The constituent quark masses are chosen to be $m_Q=300$ MeV
and $m_S=450$ MeV.

We vary the hadronization temperature in the region $T_{had}=130 - 260$ MeV
and the initial transverse flow $v_T^0 = 0 - 0.7$. The obtained
transverse spectra for the $\phi$ and the $\Omega$
are fitted in the measured transverse momentum region 
$0.5 < m_T-m_\phi < 2.2$ GeV for $\phi$ and 
$0.3 < m_T-m_\Omega < 1.5$ GeV for $\Omega$, following the procedure of
the NA50 and WA97 Collaboration,
respectively. We compare the theoretical slopes to the
experimental data. Fig.~1 displays
the values of $\chi^2$ obtained from the measured and the
calculated  slopes of the $\phi$ and the $\Omega$ spectra.
The area inside the solid contour line indicates 
$\chi^2 < 3.67$. 

One can see a very strong correlation between the
hadronization temperature and the initial transverse flow, which
can be characterized by the expression 
\begin{equation}
T_{had} + a \cdot (v_T^0)^2=0.25\  GeV \ ,
\label{tvt}
\end{equation}
where $a=0.37$ GeV.
This correlation is indicated by the solid dark line in Fig.~1.
Considering only the $\chi^2<3.67$ values, a large temperature region 
is allowed, namely $T_{had} = 160 - 230$ MeV 
(paired with the appropriate initial transverse flow $v_T^0$),
for an initial condition for the hadronization.
On the bottom part of Fig.~1 we show the details of our fit
in an enhanced way, displaying the values of $1/\chi^2 < 3$ on a lego plot.
Here an 'excellent' agreement can be seen 
between the result of the MICOR model and the 
experimental data in the region
$T_{had} = 175 \pm 15$ MeV and $v_T^0 = 0.46 \pm 0.05$.

To avoid later confusion between the theoretical 
result of this paper and the experimental data of NA50 \cite{QM99NA50}
and NA49 Collaboration \cite{NA49phi, Roland98},
Fig.~2 displays our recent information on the transverse momentum spectra
of the $\phi$ meson. In the MICOR model,  according to eq.(\ref{tvt})
we use $T_{had}=175$ MeV and $v_T^0=0.46$
to calculate the theoretical transverse momentum spectra of the $\phi$ meson.

The full squares show the NA49 data as
reconstructed from Ref.~\cite{NA49phi}. The solid line
indicates the $T=295 \pm 15$ MeV slope with the parameterization
\begin{equation}
dN/m_Tdm_T \ \propto \  \exp(-m_T/T)
\label{alpha0}
\end{equation}
in the momentum region
$0.02 < m_T-m_{\phi} < 1.5$ GeV.

The dashed line indicates the results of the
NA50 Collaboration ~\cite{QM99NA50}, namely the $T=222 \pm 6$MeV slope with
the parametrization 
\begin{equation}
dN/m_Tdm_T = C \cdot m_T \cdot K_1(m_T/T) 
\end{equation}
in the momentum region $0.5 < m_T-m_{\phi} < 2.2$ GeV. 
The dotted line, which parallels the dashed one,
indicates the same result as NA50, when using
the expansion of the Bessel function
\begin{equation}
dN/m_Tdm_T = {\tilde C} \cdot m_T^{1/2} \exp(-m_T/T) \ ,
\label{alpha12}
\end{equation}
where $ {\tilde C} = C \cdot \sqrt{T\pi/2}$.
In this momentum region the difference between
these two parametrization is negligible, and so the results of NA50
can be compared with the results of WA97 \cite{SQM98WA97}, 
who used the parametrization of eq.~(\ref{alpha12}).

Fig.~2 shows that the slopes of the NA49 Collaboration is valid
for the transverse momentum region $m_T-m_{\phi} < 1$ GeV
and the slope of the NA50 Collaboration is valid
for $m_T-m_{\phi}> 0.3$ GeV. The MICOR result was chosen to agree
with the NA50 measurement, and it agrees with the experimental 
results of the NA49 in the region $m_T-m_{\phi}> 0.3$ GeV.
In parallel, Fig.~2 demonstrates that 
the coalescence process creates
a non-thermal momentum distribution,  and the obtained
momentum distribution for the $\phi$
can  mimic a thermalized final state in large momentum region. 
(The fluctuations on the MICOR
results are coming from the applied Monte-Carlo evaluation of the
phase-space integral.)
Fig.~2 shows that the MICOR model fails to reproduce the NA49 data
in the momentum region $m_T-m_{\phi}< 0.3$ GeV, which requires 
further improvements at small $p_T$. However this disagreement will not 
influence our analysis and discussion at larger $p_T$.

We apply the obtained correlation between the hadronization temperature
and early transverse flow from eq.(\ref{tvt}) and
recalculate the slopes of the transverse momentum spectra
for $\Omega$ and $\phi$. Fig.~3. shows our results, displaying the
{\it "Data/Theory"} ratios for the different
$T_{slope}$ values (including the experimental error bars)
as a function of hadronization
temperature, $T_{had}$. This figure confirms the validity of the
function used in eq.(\ref{tvt}).

We also calculate the $\rho/\omega$ spectra.
In the MICOR model the $\rho/\omega$ particle is formed 
in one step from  
the coalescence  of a quark and an antiquark. The theoretical
value of this slope parameter was fit in the 
measured transverse momentum region $1.5 < m_T < 3.2$ and compared
to the experimental value $T(\rho)=219 \pm 5$ MeV.
The bottom part of Fig.~3 shows that the
pairs of $T_{had}$ and $v_T^0$ parameters which satisfy eq.(\ref{tvt})
by reproducing the slope of $\phi$ and $\Omega$, also reproduce
the measured slope of $\rho$. This occurs especially in the
region $T_{had}=175 \pm 15$ MeV.

This result is very surprising. We could expect that the weakly interacting
$\phi$ and $\Omega$ will conserve their transverse momentum distribution
from the  initial hot phase (according to our assumption 
this is a strongly
coupled deconfined phase). But the $\rho$ particle is very strongly 
coupled to nucleons and pions \cite{rho}, thus final state interactions should
modify their spectra with an extra transverse boost.
It was hardly expected, that  the $\rho$ meson  could conserve any of his 
early transverse momentum distribution. 

With this surprising result, we collect the available data on the 
$T_{slope}$ extracted by the parametrization of eq.~(\ref{alpha12}),
similarly to $\rho$, $\phi$ and $\Omega$. 
We show these experimental data on Fig.~4.
The data on $K^0_S$, $\Lambda$, ${\overline \Lambda}$, $\Xi^-$, 
${\overline \Xi}^+$ and $\Omega$ can be found in Ref.~\cite{SQM98WA97}.
The data on $\rho$ and $\phi$ are from Ref.~\cite{QM99NA50}.
For $\pi^+$, $\pi^-$, $K^+$, $K^-$, $p^+$ and ${\overline p}^-$
we considered the NA44 data \cite{NA44prldata} and 
using eq.~(\ref{alpha12}) fit these
data in the momentum region $m_T-m_i > 0.3$ GeV.  We obtained 
$T(\pi^+)=150$ MeV,
$T(\pi^-)=145$ MeV,
$T(K^+)=225$ MeV,
$T(K^-)=216$ MeV,
$T(p^+)=234$ MeV,
$T({\overline p}^-)=223$ MeV.
These data followed the tendency that the value of the $T_{slope}$
decreases with 25-30 MeV when using eq.~(\ref{alpha12}) as opposed to
eq.~(\ref{alpha0}), see e.g. Ref.~\cite{SQM98WA97}.
In Fig.~4, the dotted line was drawn across the $\rho$, $\phi$ and
$\Omega$. 

Fig.~4 helps to estimate the efficiency of the final state interactions
in the transverse momentum region $m_T-m_i > 0.3$ GeV.
It reveals that the separation of the $p^+$, ${\overline p}^-$, $\Lambda$,
${\overline \Lambda}$, $\Xi^-$ and ${\overline \Xi}^+$
from the weakly interacting $\phi$ and $\Omega$ is much
smaller than indicated in Refs.~\cite{Roland98,Grabler99}.
Even more, according to our fit, the slope of $p^+$ and ${\overline p}^-$
are very close to that of the weakly interacting $\phi$ meson. 
Thus the slope of the $\rho$ meson and its restoration in the MICOR
model is not surprising anymore.
Fig.~4 indicates that in the transverse momentum region
$m_T-m_i > 0.3$ GeV the hadronic spectra may not suffer large modifications
because of secondary collisions. Pions could be an exemptions, 
since their separation
from the other particles may indicate extra-long hadronic evolution, or their
appearance from resonance decays.
These results demand further investigation into the
properties of the hadronic spectra at $m_T-m_i < 0.3$ GeV.
\bigskip

In this paper, we analyzed the recent experimental results 
on the production of the $\phi$ meson and the $\Omega$ baryon,
using the MICOR model.
We have found a strong correlation between the hadronization temperature 
and the transverse flow of the early deconfined state, 
where the $\phi$ and $\Omega$ were created from quark-antiquark plasma.
We have found that $T_{had}=175 \pm 15 $ MeV and $v_T^0=0.46 \pm 0.05$
is favoured by data.  In this temperature region the MICOR model 
also reproduced the measured $\rho/\omega$ slope.
This result indicates that final state interactions modify very weakly
the transverse momentum spectra of the $\rho/\omega$ meson in the 
transverse momentum region $1.5 < m_T < 3.2 $ GeV.
In the large $m_T$ region the final state interactions may not modify the
transverse slopes created in an early state.
Further calculations are needed to confirm if all hadronic slopes can be 
reproduced by the MICOR model which assumes quark coalescence.
A positive result may be interpreted as the formation of a deconfined phase
(namely a massive quark-antiquark plasma)
in the early stage of the PbPb collision at CERN SPS.
\bigskip

{\bf Acknowledgment:}
We thank T.S. Bir\'o, M. Gyulassy, B. M\"uller, N. Xu and 
J. Zim\'anyi for stimulating discussions. We thank
N. Xu to make available the published NA44 data. 
One of the author (P.L.) is especially grateful to S. Vance for 
his comments and his careful reading of the manuscript.
This work was supported by the OTKA Grant No. T025579,
the US-Hungarian Joint Fund No. 652, 
and partly by the DOE Research Grant under Contract No.
de-FG-02-93ER-40764.

\newpage

\begin{center}
\vspace*{18.0cm}
\includegraphics{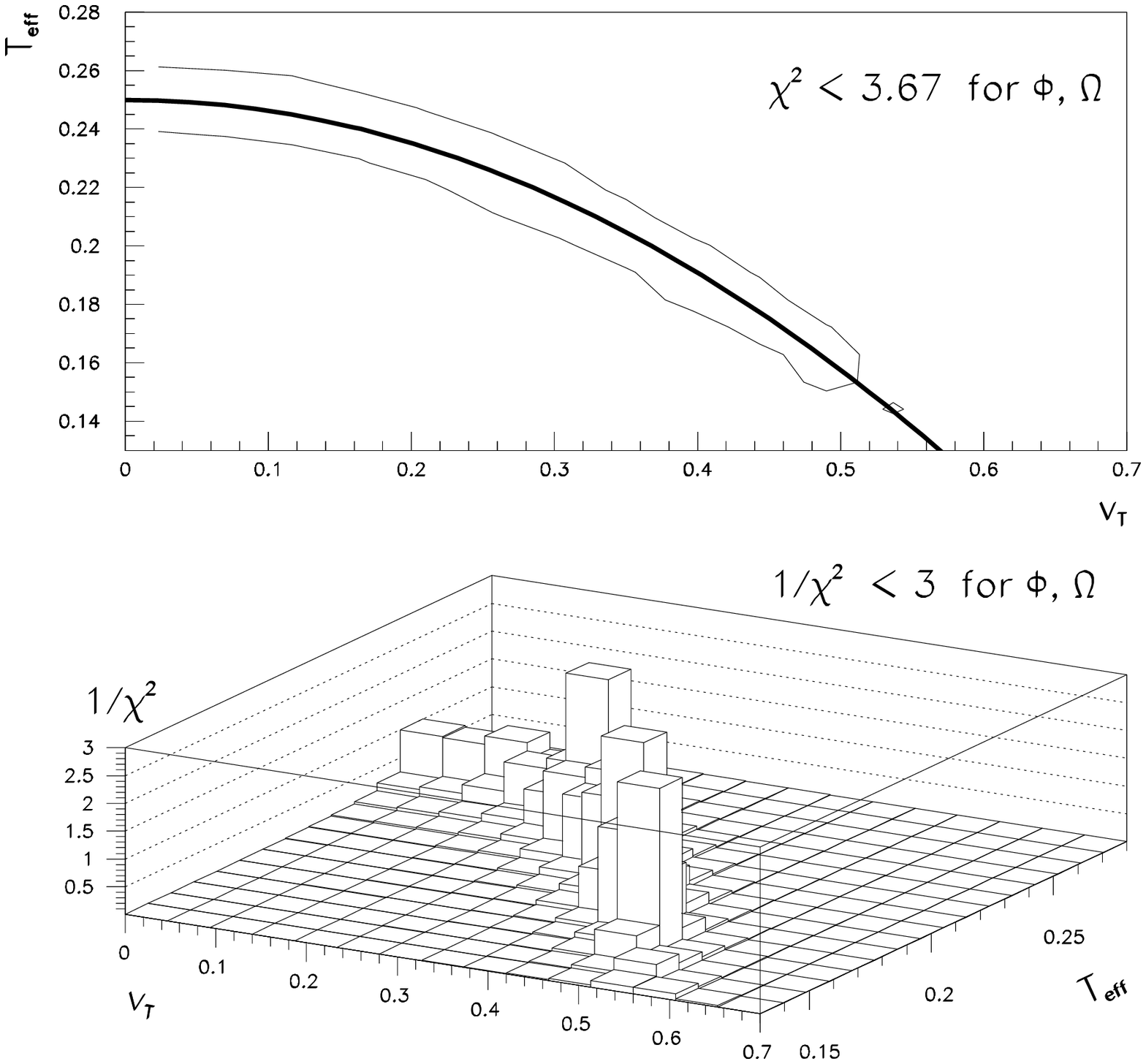}
\begin{minipage}[t]{15.cm}
{ {\bf Fig.~1.}
The $\chi^2$-fit of the calculated $T_{eff}$ slopes for 
the $\phi$ meson and the $\Omega$ baryon.
The area inside the solid contour line indicates $\chi^2<3.67$,
The solid dark line shows the fit
$T_{had} + a \cdot (v_T^0)^2=0.25$ GeV, where $a=0.37$ GeV.
On the bottom figure we show our fit
in an enhanced way, displaying the values of $1/\chi^2 < 3$ on a lego plot.
}
\end{minipage}
\end{center}

\newpage

\begin{center}
\vspace*{18.0cm}
\includegraphics{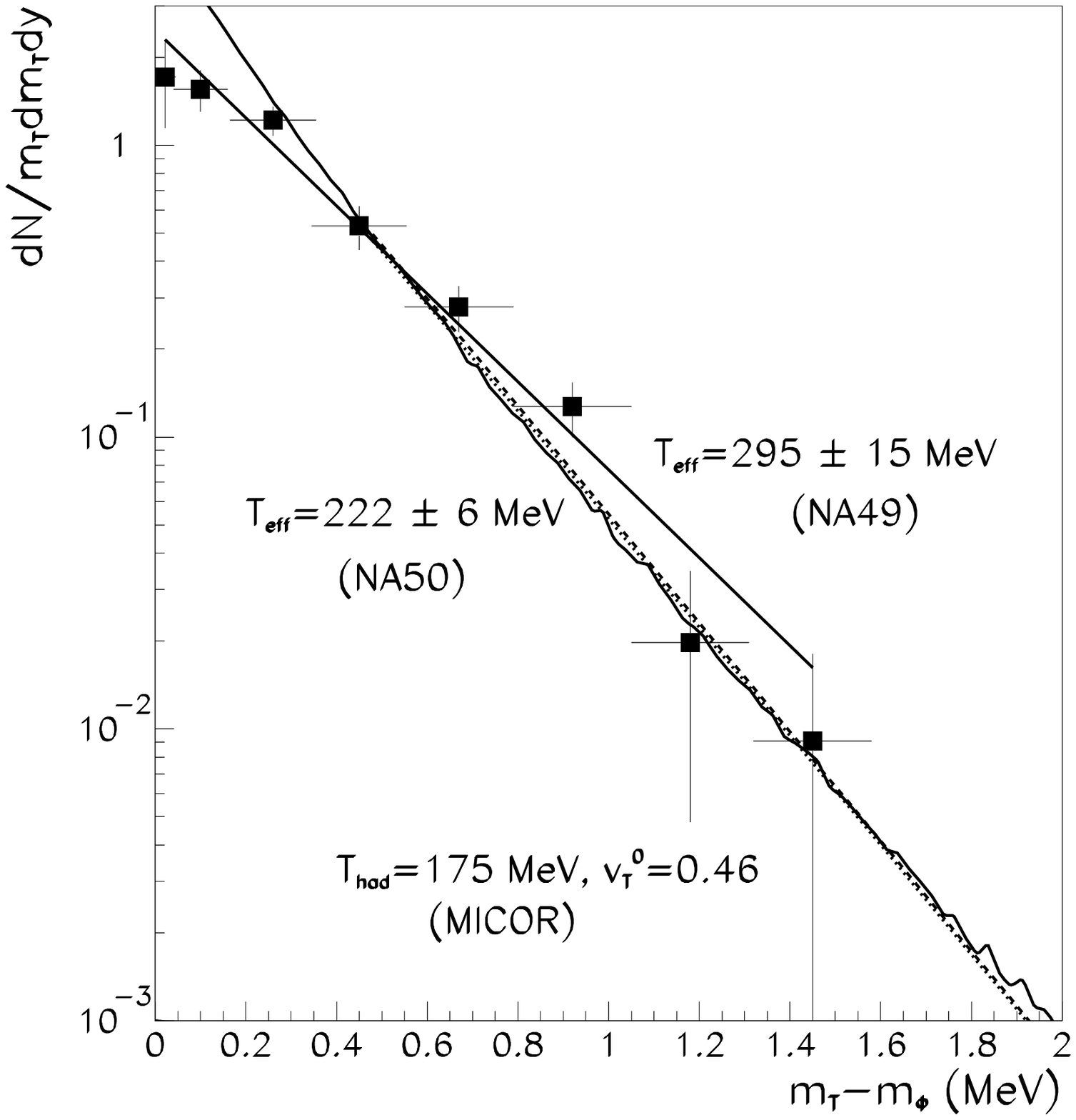}
\begin{minipage}[t]{15.cm}
{ {\bf Fig.~2.}
The transverse momentum spectra of the $\phi$ 
meson measured by the NA49 Collaboration
in Pb+Pb collision at 158 AGeV (full squares) \cite{NA49phi}. 
The solid dark line indicates the
effective slope $T=295 \pm 15$ MeV published by NA49.
The dashed line (and the overlapping dotted line) indicates
the measured slope of the NA50 Collaboration \cite{QM99NA50}
in the transverse momentum region
$0.5 < m_T - m_\phi < 2.2$ (for explanation see text).
The thin full curve with fluctuations is the MICOR result for $\phi$ production
calculated at $T_{had}=175$ MeV and $v_T^0=0.46$.
}
\end{minipage}
\end{center}

\newpage
\begin{center}
\vspace*{18.0cm}
\includegraphics{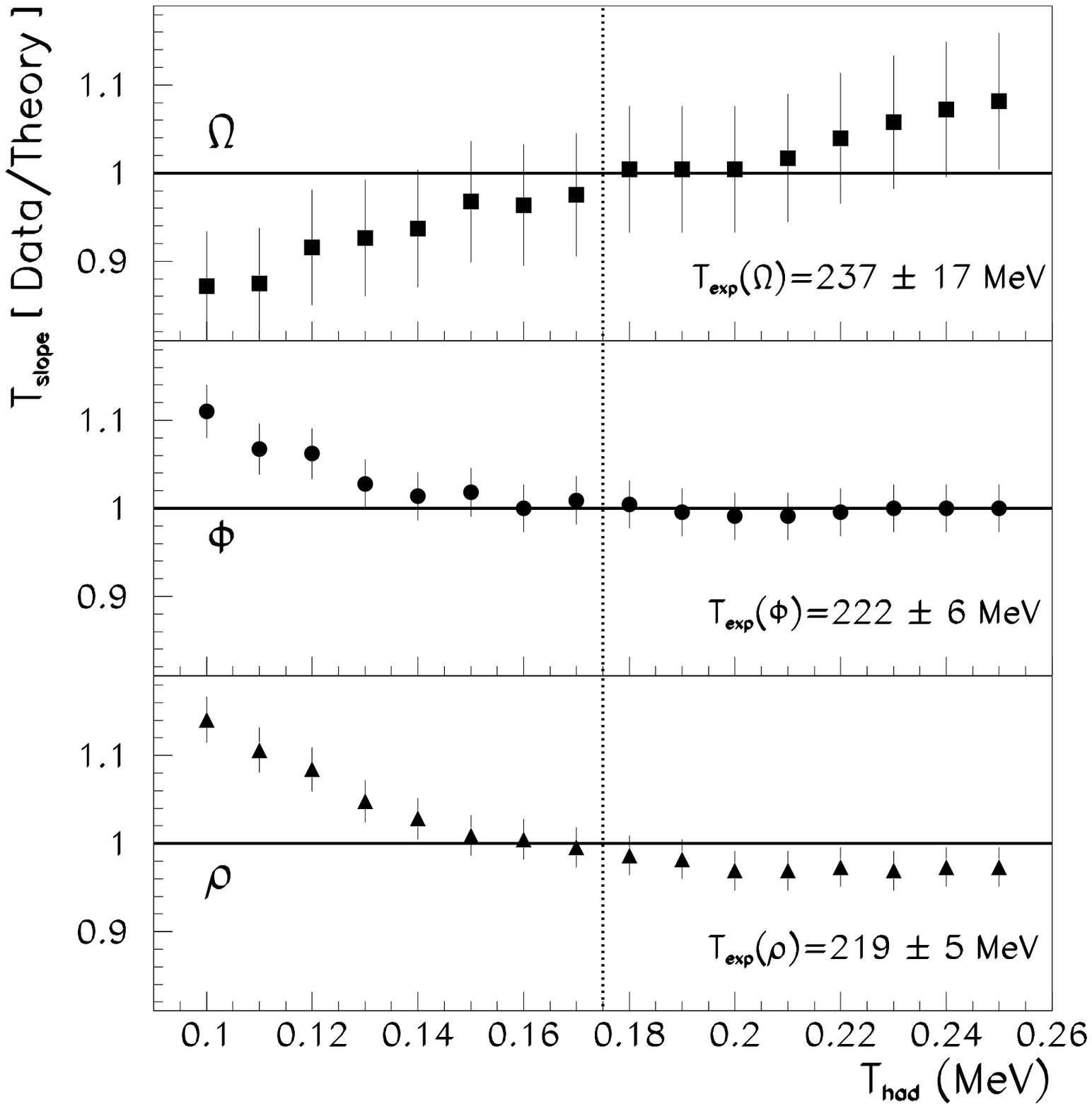}
\begin{minipage}[t]{15.cm}
{ {\bf Fig.~3.}
The {\it "Data/Theory"} values for 
recalculated $T_{slope}(\Omega)$ and $T_{slope}(\phi)$   
as a function of hadronization temperature $T_{had}$
with transverse flow $v_T^0$ satisfying eq.(\ref{tvt}).
We displayed the same ratio for the $\rho$ particle 
calculated in the MICOR model (bottom part).
Vertical dotted line indicates $T_{had}=175$ MeV.
}
\end{minipage}
\end{center}

\newpage

\bigskip
\begin{center}
\vspace*{18.0cm}
\includegraphics{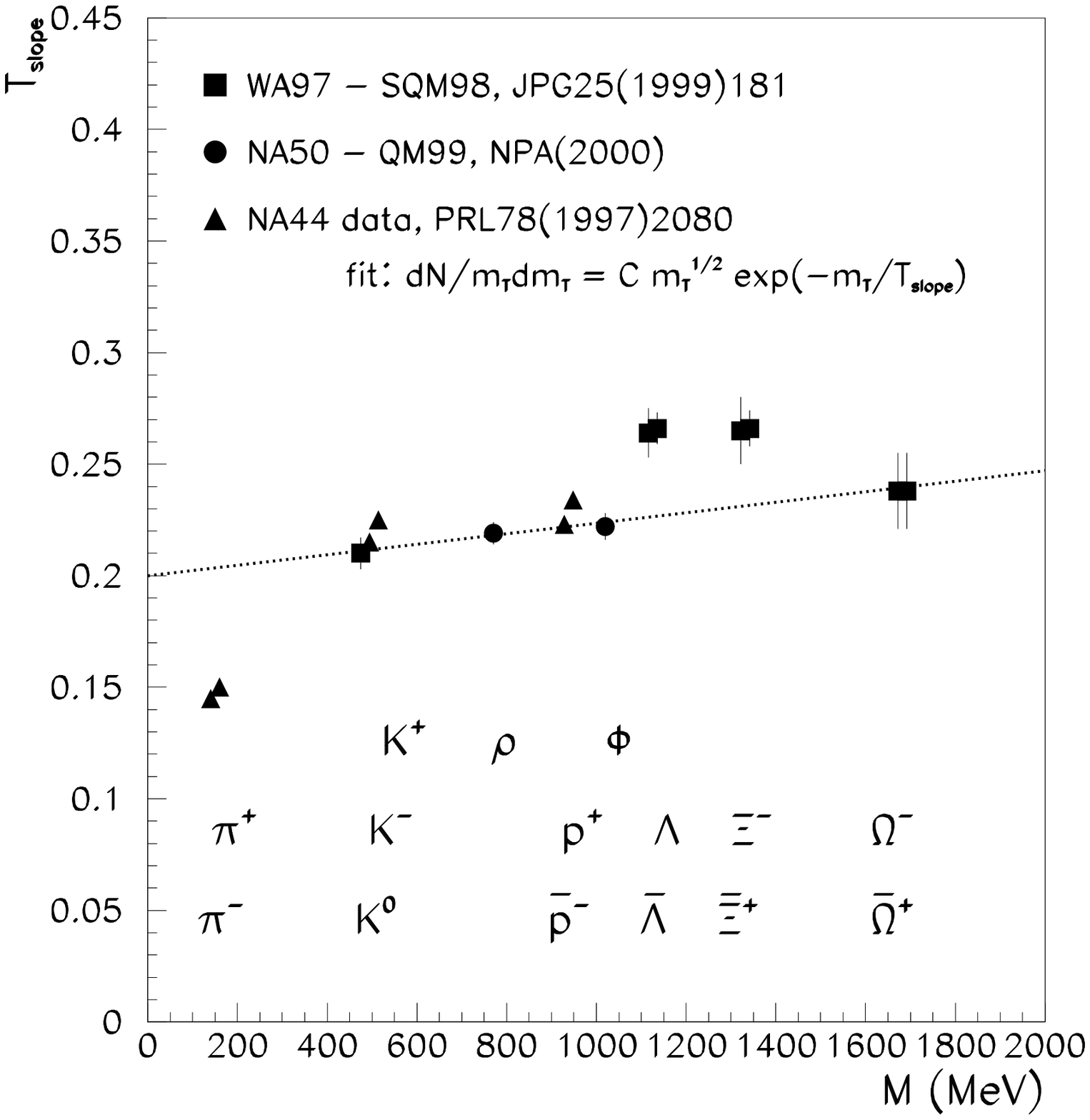}
\begin{minipage}[t]{15.cm}
{ {\bf Fig.~4.}
Experimental hadronic slopes of the transverse momentum spectra 
in the PbPb collision at 158 AGeV energy from WA97 \cite{SQM98WA97} (squares),
NA50 \cite{QM99NA50} (dots) and NA44 Collaboration \cite{NA44prl}
(triangulars).
The results of WA97 and NA50 were fitted originally by eq.(\ref{alpha12})
and we fit the NA44 data on $\pi^\pm$,
$K^\pm$, $p^+$ and ${\overline p}^-$ \cite{NA44prldata} 
in the same way in the momentum region
$m_T-m_i > 0.3$ GeV.
}
\end{minipage}
\end{center}


\begin{thebibliography}{99}
\bibitem{K1Bes}
These slope parameter were obtained in Ref.~\cite{QM99NA50}
with fitting the transverse  momentum
distribution by 
$dN/dm_T \propto m_T^2~K_1(m_T/T)$.
\bibitem{Ex32}
This slope parameter was obtained in Ref.~\cite{SQM98WA97}
with fitting the transverse momentum distribution by
$dN/dm_T \propto m_T^{3/2} \exp(-m_T/T)$.
This is close to be identical with the fit in Ref.~ \cite{K1Bes}
in the applied $m_T$ region.
\end{thebibliography}
\end{document}